\documentclass[aps,twocolumn,prb,superscriptaddress,longbibliography,reprint]{revtex4-1}
\usepackage{amsfonts}
\usepackage{mathrsfs}
\usepackage{amsmath}
\usepackage{color}
\usepackage{graphicx}
\usepackage{bm}
\usepackage{amssymb}
\usepackage{xspace}
\usepackage{epstopdf}
\usepackage{longtable}
\usepackage{multirow}
\usepackage{booktabs}
\usepackage[colorlinks=true, letterpaper=true, pdfstartview=FitV, linkcolor=blue, citecolor=blue, urlcolor=blue]{hyperref}

\begin{document}

\title{Second-Order Real Nodal-Line Semimetal in Three-Dimensional Graphdiyne}

\author{Cong Chen}
\thanks{C. Chen and X.-T. Zeng contributed equally to this work.}
\affiliation{School of Physics, Beihang University, Beijing 100191, China}
\affiliation{Research Laboratory for Quantum Materials, Singapore University of Technology and Design, Singapore 487372, Singapore}

\author{Xu-Tao Zeng}
\thanks{C. Chen and X.-T. Zeng contributed equally to this work.}
\affiliation{School of Physics, Beihang University, Beijing 100191, China}

\author{Ziyu Chen}
\affiliation{School of Physics, Beihang University, Beijing 100191, China}

\author{Y. X. Zhao}
\affiliation{National Laboratory of Solid State Microstructures and Department of Physics, Nanjing University, Nanjing 210093, China}
\affiliation{Collaborative Innovation Center of Advanced Microstructures, Nanjing University, Nanjing 210093, China}

\author{Xian-Lei Sheng}
\email{xlsheng@buaa.edu.cn}
\affiliation{School of Physics, Beihang University, Beijing 100191, China}

\author{Shengyuan A. Yang}
\address{Research Laboratory for Quantum Materials, Singapore University of Technology and Design, Singapore 487372, Singapore}

\begin{abstract}
Real topological phases featuring real Chern numbers and second-order boundary modes have been a focus of current research, but finding their material realization remains a challenge. Here, based on first-principles calculations and theoretical analysis, we reveal the already experimentally synthesized three-dimensional (3D) graphdiyne as the first realistic example of the recently proposed second-order real nodal-line semimetal. We show that the material hosts a pair of real nodal rings, each protected by two topological charges: a real Chern number and a 1D winding number. The two charges generate distinct topological boundary modes at distinct boundaries. The real Chern number leads to a pair of hinge Fermi arcs, whereas the winding number protects a double drumhead surface bands. We develop a low-energy model for 3D graphdiyne which captures the essential topological physics. Experimental aspects and possible topological transition to a 3D real Chern insulator phase are discussed.

\end{abstract}

\maketitle

Topological states of matter have been attracting tremendous research interest. The field was initiated with the study of topological insulators~\cite{Hasan_RMP,Qi_RMP,ShunQingShen_TI,BernivigTopological} and then extended to topological semimetals~\cite{Bansil_RMP,ArmitageRMP2018,Burkov2016,Yang2016,Dai2016,ZhaoYX2013}. In a topological semimetal, the band structure possesses topologically protected nodal features at the Fermi level, such as nodal points~\cite{wan2011topological,XuG_HgCr2Se4,YoungKane2012,WangZJ2012,WangZJ2013}, nodal lines~\cite{SAY2014PRL,WengHM2015PRB,YuRui2015,Kim2015PRL,fang2016topological,bian2016topological} or even nodal surfaces~\cite{Zhong2016,Liang2016,WuWK2018}, generating novel quasiparticle excitations in the bulk. In addition, like topological insulators, the bulk topology of a three-dimensional (3D) topological semimetal usually dictates protected states located on its 2D surfaces~\cite{wan2011topological,SAY2014PRL,WengHM2015PRB,Chiu2016}. This bulk-boundary correspondence was generalized with the recent development of higher-order topological insulators~\cite{ZhangFan_PRL2013,Hughes2017,Langbehn2017,SongZD2017,Hughes2017b,Schindler2018SA,Schindler2018}. A 3D higher-order topological insulator has protected states not on its surfaces but at lower-order boundaries, such as hinges or corners. It was later found that certain nodal-point semimetals may also have a higher-order topology~\cite{Hughes2018HOSM,HOWS2020,AkbarGhorashi2020,wieder2020strong,wang2020boundary}.

\begin{figure}
  \includegraphics[width=8 cm]{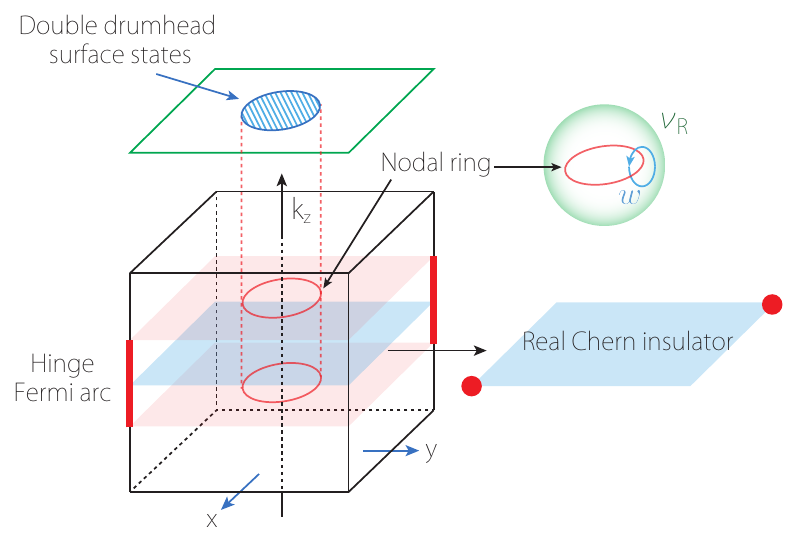}
  \caption{Schematic figure showing the second-order real nodal line semimetal phase. Here, each horizontal 2D plane between the two nodal rings is a 2D real Chern insulator with $\nu_R=1$, with a pair of topological corner states. These corner states constitute the two hinge Fermi arcs for the 3D system. Besides the nontrivial $\nu_R$, each nodal ring carries a 1D topological charge $w$, which protects the double drumhead surface bands. {Here, the nodal rings, including the inset, are shown in momentum space.}}
\label{schematic}
\end{figure}

Most recently, Wang \emph{et al.}~\cite{wang2020boundary} proposed a novel topological state---the second-order real nodal-line semimetal, protected solely by the spacetime inversion symmmetry $PT$. As illustrated in Fig.~\ref{schematic}, the state has pairs of nodal lines in the bulk, and the second-order topology is manifested in the hinge Fermi arcs located on a pair of hinges connected by $PT$. Importantly, the $PT$ symmetry enforces a real character for the band structure,
such that the bulk topology can be characterized by the $\mathbb{Z}_2$ valued real Chern number $\nu_R$ (also known as the second Stiefel-Whitney number)~\cite{ZhaoLu2017PRL,BJYang_CPB}. The bulk nodal lines and the hinge Fermi arcs are directly resulted from the nontrivial $\nu_R$ distribution.

The novel state was demonstrated by a theoretical Dirac model in Ref.~\cite{wang2020boundary}. However, like most higher-order topological states, a material realization of the state is still missing. Currently, the research on higher-order topology is severely hindered by the lack of realistic material candidates~\cite{Schindler_SciAdc,Schindler2018,Yue2019ws,WangZJ2019PRL,WanXG2019NP,XuYF2019,ZhangRX2020PRL,GDY_PRL,Liu2019b,cchen2020PRL}: the only experimentally verified second-order topological insulator material so far is Bi~\cite{Schindler2018}, {and features of combined higher-order and weak topology were reported in Bi$_2$TeI~\cite{Avraham2020}.} Thus, it is an outstanding challenge to search for suitable existing materials that realize the second-order nodal-line semimetal state.

In this work, we address this challenge by revealing 3D graphdiyne, which has already been synthesized in experiment~\cite{LiYL2010,Ruben2017,Jia2017cs,YulinagLi_ChemRev,ZhangJin_ChemSocRev}, as the first example of a second-order real nodal-line semimetal. By first-principles calculations and symmetry analysis, we show that the material hosts a pair of bulk nodal rings and hinge Fermi arcs, following the features in Fig.~\ref{schematic}, which are protected by the real Chern number. Besides the real Chern number, each ring also has an additional 1D topological charge. Interestingly, combined with the sublattice symmetry, the 1D charge leads to a pair of drumhead surface bands bounded by the projected nodal rings, distinct from usual nodal-line semimetals which only has a single drumhead surface band. We develop an effective model for 3D graphdiyne, which captures its topological features and can serve as a starting point for further studies of its properties. In addition, we point out that a tensile strain along the $c$-axis could drive the system towards a 3D real Chern insulator phase. Our work uncovers the intriguing topological character of 3D graphdiyne, and promotes it as a realistic material platform for the study of higher-order real topological phases.

\begin{figure}
  \includegraphics[width=8 cm]{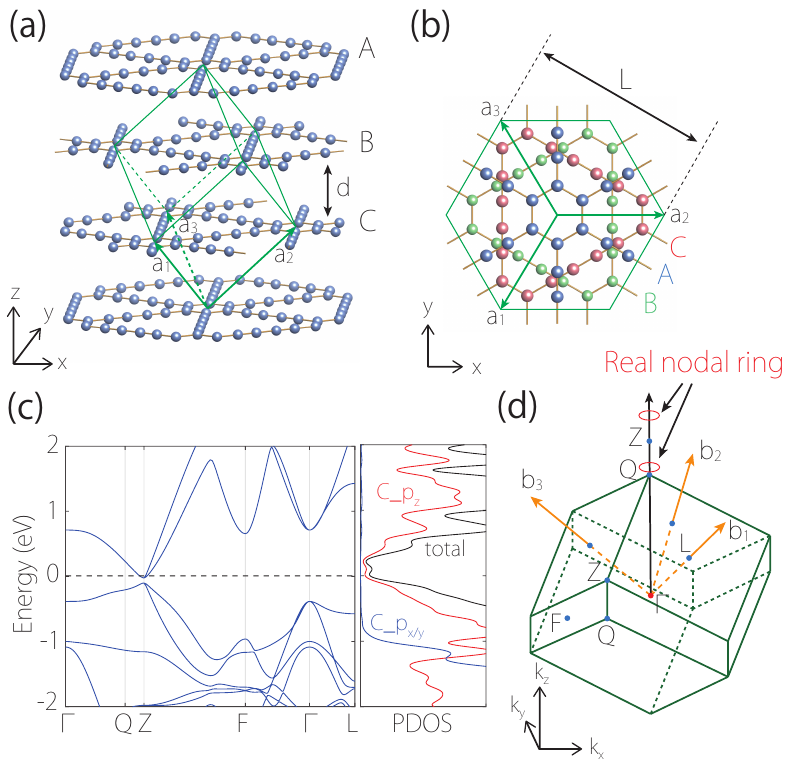}
  \caption{(a) Crystal structure of 3D graphdiyne, with three primitive lattice vectors denoted by $a_{1,2,3}$. $d$ is the interlayer distance. (b) Top view of the structure. Three layers in the unit cell are labeled with different colors. (c) Calculated band structure and projected density of states (PDOS). (d) Brillouin zone and the location of the two nodal rings. The inset shows an enlarged view around $Z$. }
  \label{band}
\end{figure}

{\color{blue}{\em Crystal structure.}}
3D graphdiyne is a famous carbon allotrope. It was first synthesized by Li \emph{et al.} in 2010 via a cross-coupling reaction method~\cite{LiYL2010}. Later on, several other methods for fabricating the material were developed, such as the explosion method~\cite{zuo2017facile,wang2018ultrafastly}, the interfacial synthesis method~\cite{interfactialGDY} and the van der Waals (vdW) epitaxy~\cite{gao2018ultrathin,zhou2018synthesis}.
The obtained 3D graphdiyne has the structure shown in Figs.~\ref{band}(a,b). It possesses a vdW layered structure, consisting of 2D graphdiyne sheets stacked along the $c$-axis, with an ABC-stacking pattern. Each graphdiyne monolayer can be viewed as being composed of benzene rings connected by diacetylenic linkages and arranged in a hexagonal lattice. The ABC-stacking pattern was found to be energetically favored and confirmed by experimental characterizations,
such as electron diffraction, high-resolution TEM and x-ray diffraction~\cite{li2018direct,interfactialGDY,gao2018ultrathin,zhou2018synthesis}.

The structure belongs to the rhombohedral crystal system, with space group No.~166 ($R \overline{3} m$). The stacking preserves the $C_{3z}$ symmetry, but breaks $C_{2z}$ which exists for a monolayer. Importantly, both the inversion $P$ and the time reversal $T$ symmetries are preserved in 3D graphdiyne. Meanwhile, the system is essentially spinless, since the spin-orbit coupling in carbon allotropes can be neglected. These fulfil the conditions required for the real topological semimetal phase~\cite{wang2020boundary}.

From our first-principles calculation (details are presented in the Supplemental Material~\cite{SM}), the optimized lattice parameters are $L=9.46$ \AA ~and $d=3.29$ \AA, which agree well with the experimental values~\cite{interfactialGDY,gao2018ultrathin}. In calculating the band structure, we adopt the primitive cell as shown in Figs.~\ref{band}(a,b), which has a rhombohedral shape and contains 18 carbon atoms.

{\color{blue}{\em Bulk band structure.}} Figure~\ref{band}(c) shows the bulk band structure of 3D graphdiyne together with the projected density of states (PDOS) obtained from our calculations. One observes that the bands  maintain a nonzero local gap along the high-symmetry paths. At the $\Gamma$ point, the local gap is $\sim$0.96 eV, comparable with the gap size of $\sim$0.51 eV for monolayer graphdiyne. It hence seems that the system is a small gap semiconductor. However, a careful scan for the low-energy bands around the $Z$ point reveals that the conduction and valence bands actually cross at a pair of nodal rings [see Fig.~\ref{band}(d)], which agrees with previous calculations~\cite{Nomura2018PRM,ahn2018band}. Thus,  3D graphdiyne is a nodal-line semimetal, consistent with the character manifested by the PDOS plot. In Fig.~\ref{band}(d), we illustrate the two rings around the equivalent $Z$ point on the $k_z$ axis at $k_z=\pi/d$, where $d$ is the interlayer distance indicated in Fig.~\ref{band}(a). (Note that the period along $k_z$ is $2\pi/d$ in the $k$ space.)

\begin{table}[b]
  \caption{Parity information at the eight TRIM points. The coordinates of these points are given by $Z$ (0.5, 0.5, 0.5), $L_1$ (0.5, 0, 0), $L_2$ (0, 0.5, 0), $L_3$ (0, 0, 0.5), $F_1$ (0.5, 0.5, 0), $F_2$ (0, 0.5, 0.5), and $F_3$ (0.5, 0, 0.5). $n_{+}$ ($n_{-}$) denotes the number of occupied bands with positive (negative) parity eigenvalues. The real Chern number $\nu_R$ for $k_z=0$ and $k_z=\pi$ plane is 1 and 0, respectively.}
\setlength{\tabcolsep}{2.5mm}{
  \begin{tabular}{ccccclcccc}
  \hline \hline
  \multirow{2}{*}{{~}} & \multicolumn{4}{c}{$k_z=0$}             &  & \multicolumn{4}{c}{$k_z=\pi$} \\ \cline{2-5} \cline{7-10}
                                       & $\Gamma$  & $L_1$  & $L_2$  & \multicolumn{1}{l}{$F_1$} &  & $Z$    & $F_2$    & $L_3$    & $F_3$    \\ \hline
        $n_{ +}$                       & 20 & 18 & 18 & 18                    &  & 18   & 18   & 18   & 18   \\
        $n_{-}$                        & 16 & 18 & 18 & 18                    &  & 18   & 18   & 18   & 18   \\ \hline 
        $\nu_R$ & \multicolumn{4}{c}{1}     &  & \multicolumn{4}{c}{0} \\ \hline \hline
  \end{tabular}}
  \label{TRIM}
\end{table}


The nodal pattern in Fig.~\ref{band}(d) is similar to that in the schematic Fig.~\ref{schematic} for real topological nodal-line semimetals. To confirm the topological character, we evaluate the real Chern number $\nu_R$ for horizontal 2D planes on the two sides of a nodal ring. Particularly, it is convenient to do the evaluation for the $k_z=0$ and $k_z=\pi/d$ planes [see Fig.~\ref{band}(d)], because for these two planes, $\nu_R$ can be readily extracted from the parity eigenvalues at the time-reversal invariant momentum (TRIM) points on each plane, with~\cite{ahn2018band,FangChen2015PRB}
\begin{equation}
  (-1)^{\nu_R}=\prod_{i} (-1)^{\left\lfloor (n^{\Gamma_i}_-/2)\right\rfloor},
\end{equation}
where $\lfloor\cdots\rfloor$ is the floor function and $n^{\Gamma_i}_-$ is the number of occupied bands with negative parity eigenvalue at TRIM point $\Gamma_i$. The calculated values of $n^{\Gamma_i}_-$ are listed in Table~\ref{TRIM} for all the TRIM points, from which one finds that the real Chern number is nontrivial $(\nu_R=1)$ for $k_z=0$ plane, and it is trivial $(\nu_R=0)$ for $k_z=\pi/d$ plane~\footnote{$\nu_R$ may also be evaluated using the Wilson loop method, as verified in Supplemental Material~\cite{SM}.}. This confirms the physical picture in Fig.~\ref{schematic}. The switch in topology with $k_z$ from $\nu_R=1$ to $\nu_R=0$ dictates the existence of nodal lines between the two planes. It also follows that each ring carries this nontrivial 2D topological charge $\nu_R=1$ defined on a small sphere $S^2$ surrounding the ring.~\footnote{ $\nu_R$ has a compact integral expression when there are two occupied bands (see Eq.(7) in Ref.~\cite{ZhaoLu2017PRL}). However, for more than two bands, there is no such integral expression available.}

In addition, we note that each ring features another 1D topological charge, corresponding to the quantized $\pi$ Berry phase
\begin{equation}\label{w}
  w=\frac{1}{\pi}\oint_C \text{Tr} \bm{\mathcal{A}}(\bm k)\cdot d\bm k\ \mod 2,
\end{equation}
where $\bm{\mathcal{A}}$ is the Berry connection for the occupied bands and $C$ is a closed path encircling the ring. Hence, the real nodal rings in 3D graphdiyne have two $\mathbb{Z}_2$ valued topological charges $(\nu_R, w)=(1,1)$. Moreover, we note that 3D graphdiyne also possesses a sublattice (chiral) symmetry $S$, i.e., its lattice sites can be divided into two sublattices connected by the $PT$ operation. With the $S$ symmetry, the 1D topological classification can be promoted from $\mathbb{Z}_2$ to $\mathbb{Z}$~\cite{Schnyder2008PRB}, with the 1D charge $w$ replaced by $w_S$ which is just Eq.~(\ref{w}) before taking mod 2.


{\color{blue}{\em Topological surface and hinge states.}}
Distinct from previously known topological semimetal materials which has protected states only at fixed order boundaries, 3D graphdiyne has topological states at both the surface and the hinge. This is a direct result of its bulk having two  nontrivial topological invariants.

\begin{figure}
  \includegraphics[width=8 cm]{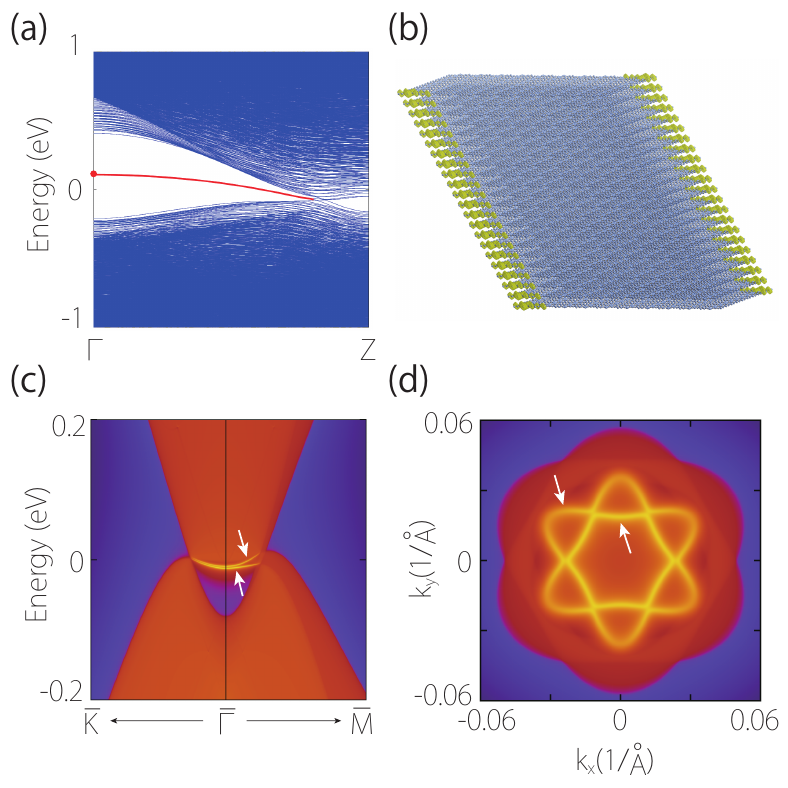}
  \caption{(a) Spectrum for a sample with tube-like geometry as illustrated in (b). The red color indicates the topological hinge band. The spatial distribution for the state marked by the red dot is shown in (b).
  (c) Projected spectrum for the (001) surface. (d) shows the constant energy slice at $-9.5$ meV. The arrows in (c) and (d) indicate the double drumhead surface bands.}
\label{surf}
\end{figure}

The 2D invariant $\nu_R$ leads to protected hinge states on a pair of $PT$ related vertical hinges, as illustrated in Fig.~\ref{schematic}. To understand this, note that each slice with a given $k_z$ in the BZ can be regarded as a 2D subsystem. The slices with $k_z\in (-k_R, k_R)$ carry nontrivial $\nu_R=1$, whereas those outside of this interval have $\nu_R=0$, where $k_R\simeq 0.864 \pi/d$ is the $k_z$ for the location of the rings in Fig.~\ref{band}(d). In other words, each slice in the region $(-k_R, k_R)$ bounded by the two rings is a 2D real Chern insulator, which must have protected zero-modes at a pair of $PT$ connected corners. And the corner zero-modes from all these nontrivial slices constitute the hinge Fermi arcs for the 3D graphdiyne. This is explicitly confirmed by our DFT calculations. In Fig.~\ref{surf}(a), we plot the spectrum for the 3D graphdiyne sample with a tube-like geometry shown in Fig.~\ref{surf}(b). One observes that there is an isolated hinge band located around zero energy in the gap, which connects the projections of the bulk nodal rings at $k_z=\pm k_R$. The hinge character is further verified by checking the spatial distribution for the states on this band, as illustrated in Fig.~\ref{surf}(b). These hinge states are the fingerprint for a second-order topological phase.

The 1D invariant on the other hand gives rise to protected surface states on the top and bottom surfaces of a sample.
Consider the straight path along the $k_z$-axis traversing the BZ. As can be explicitly seen from the effective model [see Eq.~\eqref{SONLSM} below], this 1D subsystem has the chiral symmetry and the integer valued topological charge $w_S=2$.
This means there will be a pair of drumhead surface bands bounded by the projection of the nodal rings on the top (bottom) surface. This is distinct from conventional nodal-ring semimetals, which typically only have a single drumhead surface band~\cite{SAY2014PRL,WengHM2015PRB}. Our analysis is confirmed by the DFT result in Figs.~\ref{surf}(c,d), which plots the projected spectrum for the top (001) surface.

{\color{blue}{\em Effective model for 3D graphdiyne.}}
To further understand the topological character, we develop a simple effective model for 3D graphdiyne. As the material is of vdW layered type, we adopt a layer construction approach and start by first constructing a model for a single graphdiyne layer. We note that single layer graphdiyne  has a direct band gap at $\Gamma$, and the low-energy band structure involves four bands. We impose the $D_{3d}$ point group symmetry, which is the one shared by the 3D graphdiyne bulk structure. The group contains the generators $C_{3z}$, $M_x$, and $P$, which are represented in the four-state basis at $\Gamma$ as
\begin{equation}
C_{3 z}=\sigma_0 e^{i(2\pi / 3) \tau_{z}} ,\  M_x = \sigma_z \tau_x,\ P = \sigma_z \tau_0,
\end{equation}
where $\sigma_i$ and $\tau_i$ are two sets of Pauli matrices, $\sigma_0$ and $\tau_0$ are $2\times 2$ identity matrix. In addition, we have time reversal symmetry $T = -\sigma_z \tau_x \mathcal{K}$, where $\mathcal{K}$ is the complex conjugation.
Constrained by these symmetries, the effective model for a single layer in the 3D graphdiyne up to $k^2$ order is~\footnote{Here, possible terms $\propto \sigma_0\tau_0$ is dropped as they do not affect the essential topology.}
\begin{equation}\label{Heff2D}
\begin{split}
\mathcal{H}_\text{2D}=&\left(-\Delta+m {k}_x^{2}+mk_y^2\right) \sigma_{z} \tau_{0}+v k_{x} \sigma_{x} \tau_{x}+v k_{y} \sigma_{x} \tau_{y} \\&+ \lambda\left(k_{x}^{2}-k_{y}^{2}\right) \sigma_{z} \tau_{x}+2 \lambda k_{x} k_{y} \sigma_{z} \tau_{y}.
\end{split}
\end{equation}
Here, $\Delta (>0)$ is the band gap at $\Gamma$, $m$, $v$, and $\lambda$ are real model parameters, and $m$ must satisfy the condition $m>0$ in accord with the double band inversion at $\Gamma$.  We stress that the $k$-quadratic terms in model (\ref{Heff2D}) (i.e., terms with $m$ and $\lambda$) are indispensable to
capture the correct band topology, namely, $\mathcal{H}_\text{2D}$ should be a 2D real Chern insulator with $\nu_R=1$ and protected corner zero-modes for a finite sample with $PT$ symmetry~\cite{GDY_PRL}. Without these terms, the model only describes a trivial insulator~\cite{Nomura2018PRM}.

\begin{figure}
  \includegraphics[width=8 cm]{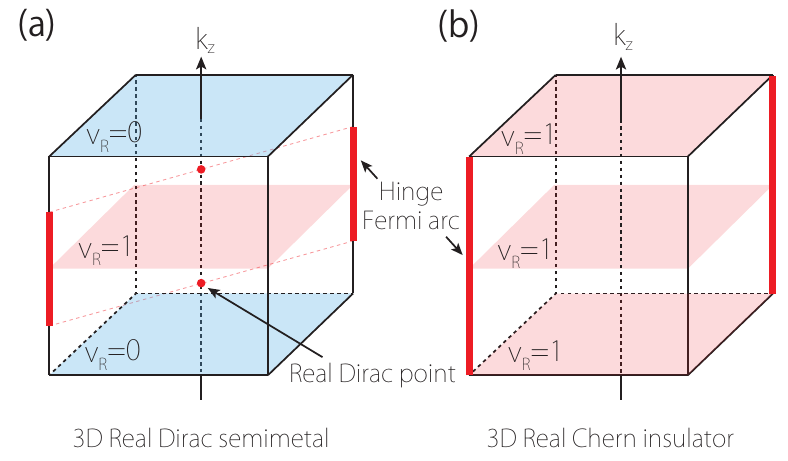}
  \caption{(a) Schematic figure showing the case with real Dirac points.  (b) shows the 3D real Chern insulator phase, where every 2D horizontal slice has a nontrivial real Chern number. }
\label{other}
\end{figure}

Next, we stack the $\mathcal{H}_\text{2D}$ layers along $z$  and add the interlayer coupling terms (which are again constrained by the symmetries discussed above). The resulting model for 3D graphdiyne takes the form of
\begin{equation}\label{SONLSM}
\begin{split}
\mathcal{H}= \mathcal{H}_\text{2D}
- u_1 \cos{(k_z d)}\sigma_z \tau_0 - u_2 \sin{(k_z d)} \sigma_y \tau_z.
\end{split}
\end{equation}
The interlayer coupling terms result in the band dispersion along $k_z$. More importantly, they modify the inverted band gap for different $k_z$ and drive a topological transition. For simplicity, let us first consider the $u_1$ coupling term. One can easily see that when $u_1>\Delta$, the conduction and valence bands will cross at $k_z=\pm k_D$ with $k_D=\pm \arccos \left(-\Delta / u_{1}\right)/d$, forming two fourfold real Dirac points on the $k_z$-axis. For 2D slices beyond the Dirac points, the double band inversion in $\mathcal{H}_\text{2D}$ is removed and they are topologically trivial with $\nu_R=0$. Thus, each Dirac point carries a nontrivial topological charge $\nu_R=1$~\cite{ZhaoLu2017PRL}, and the slice with Dirac point at $k_z=\pm k_D$ represents the critical state at the topological phase transition between 2D insulator phases with $\nu_R=1$ and $0$ [see Fig.~\ref{other}(a)]. Including the $u_2$ term will not affect this essential topology but only spread each Dirac point into a real nodal ring, resulting in the picture shown in Fig.~\ref{schematic}.

Our simple fourband effective model hence captures all the important topological features of 3D graphdiyne. It serves as a good starting point for further study of the physics of 3D real topological semimetal phases. We note that some previous works attempted to construct tight-binding models based solely on $p_z$ orbitals~\cite{Nomura2018PRM,npjpaper2020}. However, such models fail to capture the correct topology for both 2D and 3D graphdiyne: for 2D, they result in a trivial insulator; for 3D, the obtained $k_z$ interval with nontrivial $\nu_R$ and hinge states is wrong.

{\color{blue}{\em Discussion.}} We have revealed 3D graphdiyne as the first realistic material example of a second-order real nodal-line semimetal. We show that its bulk band structure features both 2D and 1D topological invariants corresponding to the real Chern number and the 1D winding number, respectively. And these invariants lead to two kinds of topological boundary states: the hinge Fermi arcs and the double drumhead surface bands.

These remarkable features contrast with conventional nodal-line semimetals and they give clear signals for experimental detection. The bulk nodal rings can be imaged by the angle-resolved photoemission spectroscopy (ARPES)~\cite{ArmitageRMP2018,fu2019dirac}. The double drumhead surface bands can be detected on the (001) surface by ARPES~\cite{Xu2015Sci,Borisenko2014PRL} or by surface sensitive probes such as scanning tunneling spectroscopy (STS)~\cite{zheng2016atomic}. The hinge states can be probed by STS, as was demonstrated for Bi~\cite{Schindler2018}.

Finally, we mention that by weakening the interlayer coupling, one can drive a topological phase transition. This can readily understood from the effective model. When the interlayer coupling $u_1$ and $u_2$ decreases, the two nodal rings will move towards $Z$, shrink into Dirac points, eventually annihilate with each other and open a bulk gap. As a result, every 2D slice normal to $k_z$ has a nontrivial real Chern number $\nu_R=1$, and the resulting state may be termed as a 3D real Chern insulator, as shown in Fig.~\ref{other}(b). In practice, the transition may be driven by applying a tensile strain along $z$, or by inserting trivial insulating layers (like BN) to form a vdW superlattice.

\begin{acknowledgments}
We thank D. L. Deng for helpful discussions. This work is supported by the NSFC (Grants No. 12174018, No. 12074024, No. 11774018, No. 11874201 and No. 12174181), and the Singapore MOE AcRF Tier 2 (MOE2019-T2-1-001).
	
\end{acknowledgments}

\bibliographystyle{apsrev4-1}
\bibliography{SONLSM_ref}



\begin{appendix}

\title{Supplemental Material for ``Second-Order Real Nodal-Line Semimetal in Three-Dimensional Graphdiyne"}
\renewcommand{\theequation}{A\arabic{equation}}
\setcounter{equation}{0}
\renewcommand{\thefigure}{A\arabic{figure}}
\setcounter{figure}{0}
\renewcommand{\thetable}{A\arabic{table}}
\setcounter{table}{0}


\section{First-principles calculation Methods}
The first-principles calculations were carried out based on the density-functional theory (DFT) as implemented in the Vienna \textit{ab initio} simulation package (VASP)~\cite{KressePRB1994,Kresse1996}, using the projector augmented wave method~\cite{PAW}.  The generalized gradient approximation (GGA) with Perdew-Burke-Ernzerhof (PBE)~\cite{PBE} realization was adopted for the exchange-correlation potential.  The plane-wave cutoff energy was set to 500 eV.  The Monkhorst-Pack $k$-point mesh~\cite{PhysRevB.13.5188} of size $11\times11\times 11$ was used for the BZ sampling in bulk calculations.  The crystal structure was optimized until the forces on the ions are less than 0.01 eV/\AA.  From the DFT results, the maximally localized Wannier functions (MLWF) for  C-$2s$ and C-$2p$ orbitals were constructed, based on which the ab-initio tight-binding models for the semi-infinite  systems were developed to study the surface and hinge spectra~\cite{Marzari1997,Souza2001,Wu2017,Green1984,Green1985}.

\begin{figure}[h!]
\includegraphics[width=8 cm]{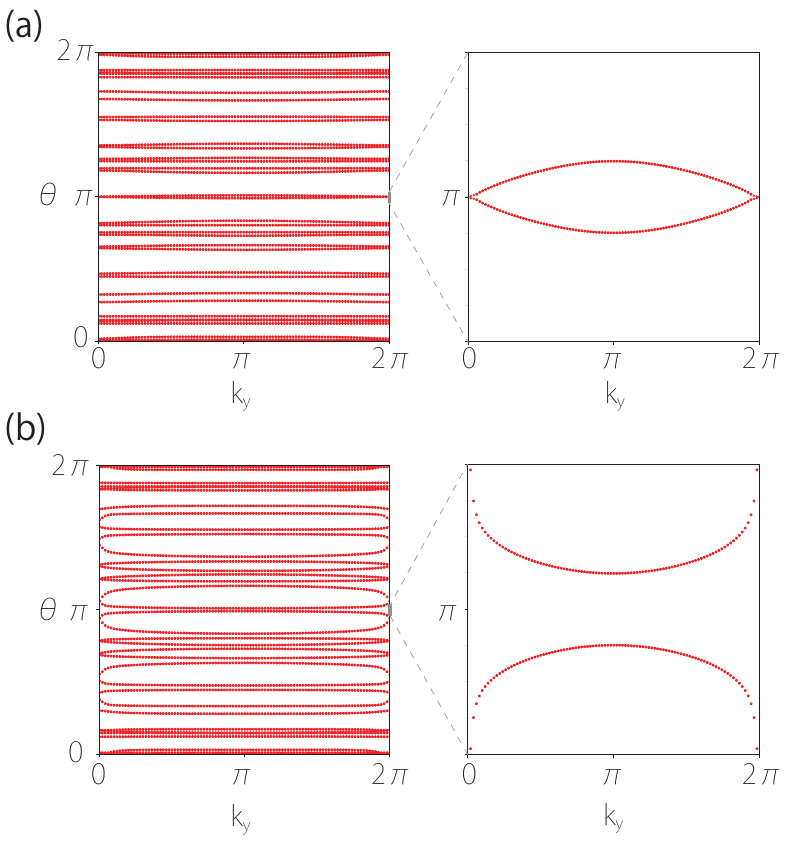}
\caption{Wilson loop spectrum calculated on (a) $k_z=0$ plane and (b) $k_z=\pi$ plane for 3D graphdiyne. The right panel shows the enlarged view near $\theta=\pi$. For $k_z=0$ ($k_z=\pi$) plane, it exhibits one (zero) crossing point with $\theta=\pi$ line, which indicates $\nu_R=1$ ($\nu_R=0$).}
\label{wilson}
\end{figure}

\section{Wilson Loop Method to verify the real Chern number}
Besides the parity analysis presented in the main text, the real Chern nmber $\nu_R$ can also be evaluated by using the Wilson-loop method~\cite{ahn2018band,chen2020graphyne}, similar to the well-known Wilson-loop method for conventional topological insulators~\cite{YuRui_PRB2011,Soluyanov2011}.
The computation will result in $N$ curves in the $\theta$-$k_y$ diagram (the Wilson loop spectrum), representing the evolution of Wannier centers, as shown in Fig.~\ref{wilson} for 3D graphdiyne.

Because of $PT$ symmetry, the Wilson loop spectrum for a 2D slice (with fixed $k_z$)  must be mirror symmetric with respect to $\theta=0$ (see Fig.~\ref{wilson}).
The real Chern number can be read off as the parity of times that the spectrum crosses the $\theta=\pi$ axis.

For $k_z=0$ plane, one observes from Fig.~\ref{wilson}(a) that there is a single crossing, hence $\nu_R=1$, i.e., the $k_z=0$ slice represents a 2D real Chern insulator. On the other hand, for the $k_z=\pi$ plane, Fig.~\ref{wilson}(b) shows there is no crossing, indicating $\nu_R=0$. These results are consistent with those obtained by the parity analysis at the TRIM points.

\section{DERIVATION OF THE EFFECTIVE MODEL }

\begin{figure}[h!]
\includegraphics[width=8 cm]{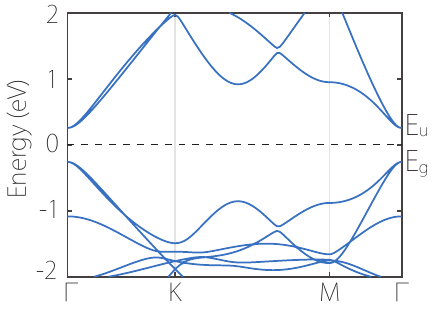}
\caption{Band structure of single layer graphdiyne. The conduction band minimum (CBM) and the valence band maximum (VBM) states have $E_u$ and $E_g$ symmetry characters, respectively.}
\label{singlelayer}
\end{figure}

Model in Eq.~(4) of the main text is for a single layer graphdiyne. Single layer graphdiyne has a direct band gap at $\Gamma$ point, and band edges at $\Gamma$ consist of two doubly degenerate pair (i.e., totally 4 states, see Fig.~\ref{singlelayer}): the valence band edge doublet corresponds to the 2D irreducible representation $E_g$ of the $D_{3d}$ group, while the conduction band edge doublet corresponds to the $E_u$ representation.  In the basis of these two doublets $(E_g, E_u)^T$, the spatial symmetry operators are hence represented by
\begin{equation}
C_{3 z}=\sigma_{0} e^{i(2 \pi / 3) \tau_{z}}, M_{x}=\sigma_{z} \tau_{x}, P=\sigma_{z} \tau_{0},
\end{equation}
where the Pauli matrices $\sigma$ and $\tau$ represent two pseudospin degrees of freedom, $\sigma$ acts on the two doublets, and $\tau$ acts on the two degenerate states within each doublet, $\sigma_0$ and $\tau_0$ are the $2\times2$ identity matrix.
Then, the form of time reversal symmetry ($T$) is constrained by requirements that $T^2=1$ (as for a spinless system) and $T$ commutes with all spatial symmetries. Under these constraints, we find its representation as $-\sigma_{z} \tau_{x} \mathcal{K}$, where $\mathcal{K}$ is the complex conjugation. Constrained by these symmetries, the effective model for a single layer in the 3D graphdiyne up to $k^2$ order is
\begin{equation}\label{Heff2D}
\begin{split}
\mathcal{H}_\text{2D}=&\left(-\Delta+m {k}_x^{2}+mk_y^2\right) \sigma_{z} \tau_{0}+v k_{x} \sigma_{x} \tau_{x}+v k_{y} \sigma_{x} \tau_{y} \\&+ \lambda\left(k_{x}^{2}-k_{y}^{2}\right) \sigma_{z} \tau_{x}+2 \lambda k_{x} k_{y} \sigma_{z} \tau_{y}.
\end{split}
\end{equation}
Here, possible terms $\propto \sigma_{0} \tau_{0}$ is dropped as they do not affect the essential topology. This is how Eq.~(4) in the main text is derived.

Next, we stack the $\mathcal{H}_\text{2D}$ layers along $z$ and add the interlayer coupling terms (which are again constrained by the symmetries discussed above). The resulting model for 3D graphdiyne takes the form of
\begin{equation}\label{SONLSM}
\begin{split}
\mathcal{H}= \mathcal{H}_\text{2D}
- u_1 \cos{(k_z d)}\sigma_z \tau_0 - u_2 \sin{(k_z d)} \sigma_y \tau_z.
\end{split}
\end{equation}

\begin{figure}[h!]
\includegraphics[width=8 cm]{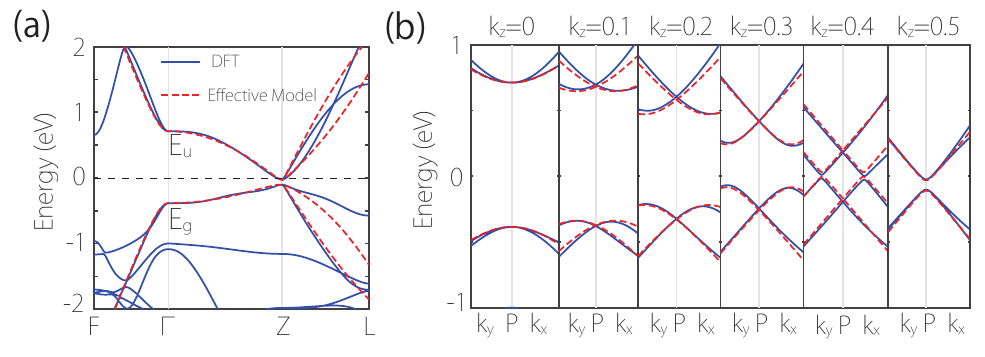}
\caption{(a) Band structure of 3D graphdiyne obtained by DFT (blue solid lines) and the fitting of effective model (red dashed lines). (b) Low-energy band structure plotted around the nodal ring. Here, point $P$ is defined by $(0, 0, k_z \cdot 2\pi/d)$. In doing the fitting, the diagonal term $(c_0+c_1 k_z^2)\sigma_0 \tau_0$ is restored. The extracted model parameters are $\Delta=0.255$~eV, $m=0.21$~eV, $v=9.5\times 10^5$~m/s, $\lambda=0.05$~eV, $u_1=0.294$~eV, $u_2=0.31$~eV, $c_0=0.164$~eV, and $c_1=-0.226$~eV. }
\label{kpDFT}
\end{figure}

The effective model captures the low-energy bands (i.e., around the Fermi energy) in first-principles calculations. And the model parameters can be extracted by fitting the DFT band structure. The result is plotted in Fig.~\ref{kpDFT} and the extracted parameters are listed in the caption. One can see that our effective model describes the low-energy bands very well.

\section{Incorrectness of tight-binding modelling with only $p_z$ orbitals}
\begin{table}[b]
  \caption{Parity information at the four TRIM points of the 2D Brillion zone for monolayer graphdiyne. The coordinates of these points are given by $\Gamma~(0,0)$, $M_1~(0.5,0)$, $M_2~(0,0.5)$ and $M_3~(0.5,0.5)$. $n_{+}~(n_{-})$  denotes the number of occupied bands with positive (negative) parity eigenvalues. The real Chern number $\nu_R$ for $p_z$ tight-binding model and DFT result is 0 and 1, respectively. }
  \setlength{\tabcolsep}{2mm}{
  \begin{tabular}{ccccclcccc}
  \hline \hline
   \multirow{2}{*}{{~}} & \multicolumn{4}{c}{$p_z$ TB Model}             &  & \multicolumn{4}{c}{DFT result} \\ \cline{2-5} \cline{7-10}
              & $\Gamma$  & $M_1$  & $M_2$  & $M_3$ &  & $\Gamma$    & $M_1$    & $M_2$    & $M_3$    \\ \hline
        $n_{+}$                       & 4 & 4 & 4 & 4                    &  & 20   & 18   & 18   & 18   \\
        $n_{-}$                       & 5 & 5 & 5 & 5                    &  & 16   & 18   & 18   & 18   \\ \hline
        $\nu_R$ & \multicolumn{4}{c}{0}     &  & \multicolumn{4}{c}{1} \\ \hline \hline
  \end{tabular}
  }
  \label{TRIM}
\end{table}

\begin{figure}[h!]
\includegraphics[width=8 cm]{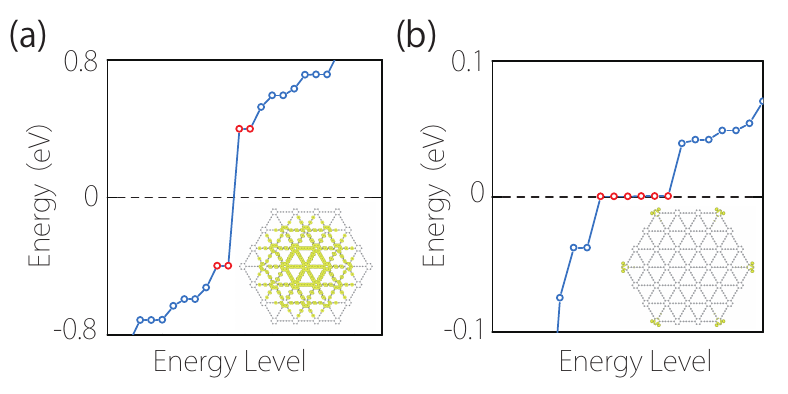}
\caption{Energy spectra of a hexagonal-shaped disk for (a) the tight-binding model including only the $p_z$ orbitals, and (b) the DFT result. The insets show the distribution of the states marked in red in the spectra. Clearly, the tight-binding model cannot capture the correct topology of graphdiyne.}
\label{2corner}
\end{figure}

As mentioned in the main text, the simple tight-binding model with only $p_z$ orbitals (denoted as $p_z$-TB model)~\cite{Nomura2018PRM,npjpaper2020} cannot capture the correct topology of both 2D and 3D graphdiyne. 
For example, in Table~\ref{TRIM}, we present the parity eigenvalue analysis obtained from the $p_z$-TB model for monolayer graphdiyne and compare it with that from our direct DFT calculations. For the $p_z$-TB model, the parity distribution for valence bands at all four TRIM points is the same. So according to the formula
\begin{equation}
(-1)^{\nu_{R}}=\prod_{i}(-1)^{\left\lfloor\left(n_{-}^{\Gamma_{i}} / 2\right)\right\rfloor},
\end{equation}
the band structure is topologically trivial, with real Chern number $\nu_R=0$. Clearly, this is wrong, since the system is nontrivial with real Chern number $\nu_R=1$, as evidenced by the full DFT result. In Fig.~\ref{2corner}, we further compare the spectra for a nanodisk. Again, the $p_z$-TB model fails to reproduce the topological corner modes as seen in the DFT result. These demonstrate that the TB model including only the $p_z$ orbitals cannot correctly capture the topology of graphdiyne.


\end{appendix}

\end{document}